\documentclass[11pt]{article}

\usepackage[final]{acl}

\usepackage{times}
\usepackage{latexsym}
\usepackage[T1]{fontenc}
\usepackage[utf8]{inputenc}
\usepackage{microtype}
\usepackage{inconsolata}
\usepackage{graphicx}

\title{What People Almost Did: \\ Evaluating LLM Social Simulations Beyond Behavioral Fit}

\author{JaeWon Kim \\
  The Information School, University of Washington \\
  \texttt{jaewonk@uw.edu} \\
  \And
  Angie Boggust \\
  MIT CSAIL \\
  }

\begin{document}
\maketitle

\begin{abstract}
LLM-based social simulations are primarily evaluated for behavioral fit, testing whether agents reproduce the actions or response distributions of the people they are simulating. However, the promise of simulation extends beyond behavioral fit. Simulations can explain human behavior, diagnose barriers, and compare large-scale interventions. These use cases depend on understanding \textit{why} people acted a certain way, not just \textit{what} they did. As a result, behavioral fit is insufficient for these types of claims because behavior underdetermines the reasoning process behind it. For instance, the behavior of staying silent may be due to disinterest or suppressed speech, and not answering a call may be due to distrust of the caller or limited phone access. In this paper, we propose \textit{representational adequacy} as a new evaluation target for LLM-based social simulations. By leveraging LLM reasoning traces, representational adequacy measures whether a simulation's scenario--reasoning--action triples preserve the reasoning process behind the behavior in a way that is faithful to the population and scenarios being simulated. We distinguish representational adequacy from interpretability and alignment metrics, propose ways to integrate it into simulation research, and pose its measurement as an open problem.
\end{abstract}

\section{Introduction}
\label{sec:intro}

Researchers simulate people when studying real populations is infeasible. Would teens share more if fewer adults could see their posts? Would more expectant mothers answer an automated health call from a voice they trusted? Running such comparisons on real populations is expensive and sometimes unethical, so simulation allows us to approximate the impact of an intervention before implementing it in real scenarios~\citep{park2025simulatinghai}.

In an LLM-based social simulation, a language model simulates a person.
It receives a persona, demographic details, memories, or interview transcripts, and acts within a scenario~\citep{park2024generative1000}. Where classical agent-based models require the researcher to write out decision rules, an LLM agent generates its own behavior and can be asked what it weighed before acting. The resulting reasoning trace may be generated before the action, stored as a structured state, elicited afterward, or reconstructed from logs. Regardless of how it is obtained, a trace represents a hypothesis about the process behind the agent's action.

Existing simulation evaluations have been based on \textit{behavioral fit}.
Behavioral fit measures whether the agent does what the corresponding person or population did. 
For instance, evaluations have scored how well agents built from long-form interviews reproduce their participants' survey responses~\citep{park2024generative1000}, how well models match group-level response distributions~\citep{hu2025simbench}, and how well agents pursue social goals in interaction~\citep{zhou2024sotopia}.

Simulations are often used not just to predict human behavior but also to explain it.
However, a person's behavior can mask their rationale.
For instance, on social media, one teenager might draft a post, worry about who will see it, and not post it, while another teen never considers posting. 
Both teens have the same behavior, but only the first teen's reasoning constitutes a social media design problem because fear of judgment suggests issues with audience boundaries and posting norms~\citep{kim2025trust, kim2025privacy}.
Alternatively, in public health settings, one pregnant woman might miss an automated maternal-health call because she shares a household phone, while another might screen the call out of distrust.
Both women miss the call, but in the first case, public health officials might consider alternative communication strategies, whereas in the second, increased public outreach is warranted~\citep{martinson2025llmmaternal}.
A simulation with high behavioral fit will reproduce not posting and missing calls, but it does not guarantee that the simulation reflects people's reasoning processes that inform interventions.

To address this gap, we argue that simulations should also be evaluated for \textit{representational adequacy}.
Representational adequacy measures whether a simulation preserves the reasoning process of the people being simulated.
To do so, we propose analyzing an LLM's scenario--reasoning--action triple to ensure consistency between the scenario an agent was placed in, the reasoning trace it produced, and the action it took.
In Section~\ref{sec:behavior} we show that behavior underdetermines reasoning in ways no behavioral benchmark can solve.
Then, in Section~\ref{sec:adequacy}, we define representational adequacy and distinguish it from other interpretability and alignment metrics. 
Finally, in Sections~\ref{sec:practice} and~\ref{sec:open}, we suggest ways that simulation research can support representational adequacy and propose its measurement as an open research problem.

\section{Behavior Underdetermines Process}
\label{sec:behavior}

Public behavior is a lossy record of why that behavior occurred~\citep{Geertz2008ThickCulturei, hedstrom2010causal}. People manage the impressions they give off~\citep{goffman1959presentation} and suppress views they believe are unpopular~\citep{noelle1974spiral, kuran1997private}.
As a result, a simulation evaluated only on the accuracy of its predicted behaviors can look right while being wrong about the explanation for its behavior~\citep{li2025simulating, puelmatouzel2026validation}.
We draw on social science literature to define a taxonomy of ways in which behavior can obscure the reasoning process behind it.

\paragraph{Missing actions.}
The relevant reasoning may never be recorded as an action.
This is a well-documented phenomenon called preference falsification, where people do not act publicly while dissenting internally~\citep{kuran1997private}.
For instance, a citizen might consider joining a protest, weigh employer retaliation, and decide to stay home.
Their lack of action in a simulation dataset is the same as a neighbor who never thought of attending. 
Yet, when using simulation to explain human behavior (e.g., to understand political approval under different policies), it is necessary to understand why people did not act.

\paragraph{Ambiguous actions.}
When an action does occur, it does not map one-to-one onto a reasoning process.
Different rationales can produce the same action (\textit{equifinality}).
For example, a junior employee might endorse a strategy because dissent feels career-threatening, while a senior colleague might find it compelling.
\citet{janis1972victims} shows how pressure toward unanimity causes people to agree despite private reservations.
The reverse (\textit{multifinality}) can also occur, in which a single reasoning process leads to different actions.
Among teens, privacy fears drive both oversharing and withdrawal~\citep{kim2025trust, kim2025privacy}.

\paragraph{Shifting composition.}
Aggregating behavior across people can obscure changes in reasoning.
A social media platform's posting rate might remain consistent across multiple years, while who posts and why they post might change.
While the behavioral distribution has not shifted, the composition of reasons beneath it has.

\paragraph{A Running Example}
Throughout the paper, we use a running example of adolescent social media use. 
Social media is a common setting for classical and LLM-based simulations~\cite{Yang2024Oasis, Lin2025SimSpark, Butler2024MisInformation, Jeon2025Simulating, Larooij2025Fix, Cheng2025Interactive, Nguyen2026Survey}.
It is susceptible to missing actions (e.g., not posting), ambiguous actions (e.g., wanting to post vs. feeling compelled to post), and shifting composition (e.g., reasons for posting changing within communities).
It is also a realistic scenario because about six in ten U.S. teens report using Instagram and roughly half report visiting it daily~\citep{pewteens2024}.
And their posting behavior is not indicative of their reasoning process, because they describe the posting environment as judgmental, performative, or not worth engaging with~\citep{landesman2024instagram, kim2024bereal, Davis2025YouEmotionsz}.

\medskip
\noindent\fbox{\parbox{0.95\columnwidth}{%
\textbf{Scenario:} A teen considers posting about having a bad day on a platform where peers, acquaintances, and a few adults can see it.\\[3pt]
\textbf{Reasoning:} Two friends are ``leaky,'' likely to screenshot or share beyond the intended audience, and on this platform, posting something vulnerable invites judgment. This outweighs the desire to share.\\[3pt]
\textbf{Action:} Delete the draft.
}}
\medskip

\section{Representational Adequacy}
\label{sec:adequacy}

LLM-based simulations offer an opportunity to not just simulate how people behave but the reasoning process behind their behavior.
Classical social simulations are based on predefined behavioral rules, so we can only evaluate the behavioral fit of a simulation. 
However, LLM-based agents can produce reasoning traces that explain their considerations alongside each action.
Current simulation benchmarks typically only retain agent actions and discard the reasoning trace.
For instance, in our running example, current benchmarks would verify that an agent simulating the scenario takes the same action as the teen (e.g., deleting the draft), but would not verify that the agent's reasoning trace aligns with the teen's reasoning.
Reasoning traces could allow us to evaluate how well simulations reflect human reasoning, and, as a result, use simulations to learn about human reasoning.

In particular, we propose that LLM-based simulations evaluate \textit{representational adequacy}.
We consider a simulation representationally adequate if the agent's reasoning trace matches what members of the represented population would say in that situation. 
To study and measure this, we propose analyzing scenario--reasoning--action triples as the unit of analysis in simulation evaluations.
Doing so ensures consistency between the scenario an agent was placed in, the reasoning trace it produced, and the action it took.

Representational adequacy is a property of a simulation's use case. A simulation is adequate for a claim to the extent that its scenario--reasoning--action triples preserve the reasoning distinctions the claim depends on. For instance, in a social medical setting, the same simulation can be adequate for studying whether fear of a leaky audience suppresses self-disclosure among teens and useless for studying platform addiction. As a result, adequacy is a function of a population (e.g., teens), a scenario (e.g., posting vulnerable content), and a claim of interest (e.g., fear suppresses self-disclosure). 

Representational adequacy differs from representational alignment.
Representational alignment measures similarity between internal states across systems~\citep{sucholutsky2023alignment}, whereas adequacy measures whether a simulation's reasoning trace matches the simulated population's reasoning.
Representational adequacy is inspired by documentation standards (e.g., datasheets, model cards, and saliency cards) that evaluate artifacts against their intended use~\citep{Mitchell2019model, gebru2021datasheets, Boggust2022SaliencyMethodsv}, and interpretability work that compares model representations against human expectations~\citep{Bolukbasi2016man, olah2018building, carter2019activation, Boggust2021SharedBehaviorv, boggust2025abstraction, pair2021what}.
A simulation is defined as adequate or inadequate for a particular claim, about a particular population, and in a particular scenario.

An important consideration for representational adequacy is how to interpret a model's reasoning process.
While LLMs can directly output a reasoning trace, work on chain-of-thought faithfulness shows that models can produce fluent traces that are causally disconnected from their answers, rationalizing after the fact and omitting the features that actually impacted their behavior~\citep{Jacovi2020Towardsz, turpin2023language, lanham2023measuring}.
Faithfulness and adequacy are distinct\,---\,faithfulness measures whether a model's reasoning reflects its computation, while adequacy measures whether model reasoning matches human reasoning and preserves the social distinctions a claim depends on.
Nevertheless, faithfulness is an important consideration when measuring representational adequacy, because for an unfaithful explanation, adequacy reports the match between the agent's rationalization of its decision rather than the underlying mechanism behind that decision.

Another important consideration is how to extract ground truth human reasoning that reflects the simulated population.
First, it should be recognizable by members of the population, such that, in our running example, it reflects the processes that teens use when considering whether to post~\citep{maxwell1992understanding}.
Second, it should align with documentation about how context, norms, and action relate in the simulated domain.
Third, it should reflect the distribution of reasoning processes across the population. 
Studying both the participant and population levels is important because reasoning is not consistent across a population. 
For instance, research has shown that posting constraint is unevenly spread across the teen population, based on judgment norms, social capital, and identity~\citep{kim2025privacy}.
A simulation that reproduces the overall silence rate while putting the constrained silence on the wrong teens misrepresents the population.
Moreover, these sources should be used together, since people's self-reported reasoning is affected by social desirability, people only have partial access to their own processes~\citep{vazire2010knows, rosenman2011measuring}, and only measuring agreement with existing literature will prevent simulation from surfacing novel considerations.

\section{Integrating with Simulation Research}
\label{sec:practice}

Representational adequacy does not replace behavioral benchmarks. 
Instead, evaluations should be based on the simulation's claims.
Behavioral fit is appropriate for a simulation used solely for prediction.
But a simulation that is used to explain, diagnose, or understand interventions should also consider representational adequacy. Doing so requires additional reporting from simulation research:

\paragraph{State the claim.}
Papers should state the type of claim they are making and for which populations and scenarios. ''Teens post less'' is a behavioral claim, while ``teens hold back because of risk rather than disinterest'' is a process claim. 

\paragraph{Match evidence to the claim.}
Research should report which evidence supports the process claim.
Simulating observed actions is evidence of behavioral fit, whereas matching individual or population reasoning supports representational adequacy.

\paragraph{Preserve the triple.}
Process claims require keeping scenario--reasoning--action triples, and collecting ground truth data (e.g., participant feedback, domain expertise, or the qualitative and empirical literature). 
This allows simulated reasoning traces to be checked against the people concerned, such as teens and social media researchers who know the barriers and the populations that experience them.

\vspace{10pt}

Reporting representational adequacy and recording scenario--reasoning--action triples gives stakeholders the agency to contest inaccurate simulations.
For example, imagine a model that matches aggregate social media nonparticipation while assigning constrained silence to the wrong groups.
A behavioral evaluation does not distinguish this, but reporting triples gives participants, reviewers, and affected communities a concrete object to contest.

Increased process understanding also creates risk.
A platform that learns which design features convert fear of judgment into self-censorship can make posting safer or make suppression more efficient.
As a result, we suggest that adequacy should pair with governance, such that where traces represent sensitive constraints, the represented communities can have a say in what is collected, how it is stored, and which uses are permitted.

\section{The Open Measurement Problem}
\label{sec:open}

By proposing a framework for representational adequacy and ways to integrate it into simulation research, we aim to encourage research to designing representational adequacy measurements.

To do so, research must first consider how to extract and verify the claims in an agent's reasoning process.
Interpretability interventions offer a starting point, from probing and causal mediation to interchange interventions and the perturbation designs in faithfulness literature~\citep{belinkov2022probing, vig2020investigating, geiger2021causal, lanham2023measuring}.
A natural adaptation is to edit individual rationales in a trace and watch whether the simulated action changes appropriately.
However, while we have focused on LLM-generated reasoning traces because they are straightforward to compute during simulation, mechanistic interpretability methods could also reveal causal concepts and circuits of agent reasoning.
Understanding the best method depends on trade-offs among faithfulness, fluency, open-weight access, and computational efficiency.

Research must also consider how to extract evidence from participants and domain experts.
While existing literature on simulated populations could provide an initial ground truth, future research is needed to develop protocols that allow stakeholders to easily provide input on appropriate reasoning processes in their domain.
This is particularly pertinent for simulations of populations without existing reference data.
Research must also consider that reasoning can be circular or contested among ground truth reasoning sources.

Meanwhile, the community can build infrastructure to support representational adequacy measures.
Benchmarks can store scenario--reasoning--action triples alongside behavioral data~\citep{hu2025simbench}.
Reviewer guidance can ask papers making process claims to report traces as evidence.
\citet{puelmatouzel2026validation} have called for related author checklists and reviewer guidelines, and representational adequacy reporting can slot directly into them.

\section{Conclusion}
\label{sec:conclusion}

Behavioral fit is an important evaluation target for LLM social simulations that predict human behavior.
However, it is not sufficient for simulations that are used to explain human behavior, because identical actions can mask underlying reasoning.
In response, we propose representational adequacy, a framework for assessing whether simulations represent decision-making processes across the simulated population.
The promise of social simulation has always been studying how people would act under conditions we cannot practically or ethically create.
Taking that promise seriously means evaluating simulations on more than what people did.
It means asking whether simulations preserve what people considered, feared, and almost did.

\bibliography{references}

\end{document}